\documentclass[reprint,superscriptaddress,amsmath,amssymb,amsart,aps,prl]{revtex4-2}
\usepackage{accents}
\usepackage{graphicx}
\usepackage{dcolumn}
\usepackage{bm}
\usepackage{braket}
\usepackage{amsthm}
\usepackage{bookmark} 
\usepackage[autostyle]{csquotes}
\usepackage{hyperref}
\graphicspath{{./figs/}}

\def\u0{\ensuremath { \hat{\bm{u}}_0}}

\newcommand{\intvol}[1]{\int \frac{{\rm d}^d  {#1}}{(2\pi\hbar)^d}}
\newcommand{\diso}{n_i v_0^2}

\begin{document}

\title{Quantum Kinetic Theory of the Linear Response \\ for Weakly Disordered Multiband Systems}

\author{T. Valet}
\email[Corresponding author :]{ tvalet@mphysx.com}
\affiliation{MPhysX O\"U, Harju maakond, Tallinn, Lasnam\"ae linnaosa, Sepapaja tn 6, 15551, Estonia}

\author{R. Raimondi}
\email[Corresponding author :]{roberto.raimondi@uniroma3.it}
\affiliation{Dipartimento di Matematica e Fisica,  Universit\`a  Roma Tre, Via della Vasca Navale 84, 00146 Roma, Italy}

\date{\today}

\begin{abstract}
A quantum kinetic theory of the linear response to an electric field is provided from a controlled expansion of the Keldysh theory at leading order, for a multiband electron system with weak scalar disorder.
The response is uniquely partitioned into intraband distribution functions and interband quantum coherences. A new explicit formula is provided for the latter, outlining their true nature as {\em local dependent} quantities, for which the mesoscopic gradients of the former is uncovered as a source term, opening a new research area. The precise connection with the Kubo formula in the ladder approximation is established. A pedagogical application to spin-orbit torque theory in the two dimensional electron gas demonstrates the striking efficiency of the quantum kinetic approach. Some important implications for the theory of orbital transport are also briefly discussed.    
\end{abstract}

\maketitle
 
{\em Introduction--} 
Linear response theory based on the Kubo formula\cite{kubo:1956,kubo:1957}, usually expressed in more practical forms of the Kubo-Bastin or Kubo-Streda types \cite{bastin:1971,streda:1975,smrcka:1977,crepieux:2001,bonbien:2020}, is instrumental to our understanding of the electron Fermi liquid properties in solids. Its application to the theory of the quantum Hall effect \cite{thouless:1982,streda:1982} provides a salient illustration of its power in this context. Berry phase effects \cite{xiao:2010} are fully included, for instance in the expressions one can derive for the intrinsic anomalous Hall effect (AHE) \cite{nagaosa:2010}, spin Hall effect (SHE) \cite{sinova:2015} or orbital Hall effect (OHE) \cite{kontani:2009,pezo:2022} conductivities. It is also naturally amenable to diagrammatic expansions \cite{abrikosov:1963,mahan:1990,rammer:2007}, allowing to account for disorder in a controlled manner. However, concrete Kubo based calculations often incorporate disorder effects only through a constant broadening approximation (CBA) \cite{ovalle:2023,go:2024}. Such an uncontrolled approximation, completely neglecting vertex corrections, is often problematic \cite{ado:2017,tang:2024}.
The Kubo formula also enables direct computations from first principles, see for instance \cite{lowitzer:2010,lowitzer:2011,freimuth:2014,bajaj:2024}, leading to quantitative predictions for actual material systems. However, in many systems under current intense investigation, such as spintronic \cite{hirohata:2020}, orbitronic \cite{jo:2024} or topological \cite{breunig:2022} nanostructured materials and devices, one is primarily interested in disorder averaged responses in finite mesoscopic samples. Full quantum simulations in such cases, specially if one wants to go beyond the CBA, become quite involved, even with the emergence of efficient real space numerical formulations \cite{groth:2014,liu:2015,fan:2021,belashchenko:2023,castro:2024}.    
Alternatively, the linear response can be derived from quasiclassical \cite{raimondi:2006}, semiclassical \cite{sinitsyn:2007,sinitsyn:2008,xiao:2017} or density matrix equation of motion \cite{culcer:2017,sekine:2017} approaches, which can be seen as attempts to incorporate leading quantum geometric and disorder corrections to the simple quasiparticle transport picture of the Boltzmann equation (BE) \cite{ziman:1960}. While this is in principle better suited to model inhomogeneous systems, these earlier works did not provide a full account of finite size effects on mesoscopic length scales, nor did they establish the exact connection to the Kubo formula at a formal level.   

In this Letter, we provide a quantum kinetic theory of the linear response to a nearly constant and uniform  electric field ${\bm E}$, for a multiband system, in presence of  weak scalar disorder. We combine  high-energy physics techniques \cite[\& references therein]{hidaka:2022}, with a gauge invariant formulation of quantum geometry \cite{graf:2021}, leading to a much streamlined, more rigorous, way of performing the semiclassical expansion of the Keldysh theory \cite{schwinger:1961,kadanoff:1962,keldysh:1965}, as detailed in the Supplemental Material \cite{supp} (see also references \cite{luttinger:1951,luttinger:1955,kohn:1959,langreth:1972,hansch:1983,larkin:1977,onoda:2006,stratonovich:1956,bialynicki:1977,vasak:1986,levanda:1994,muller:1999,levanda:2001,groenewold:1946,moyal:1949,huang:2018,lipavsky:1986,shindou:2008,wickles:2013,zhang:2020,gorini:2010} therein). Our theory uniquely partitions the response into intraband and interband contributions, uncovering mesoscopic gradients of the former as a previously overlooked source of the latter. In the uniform case, using Matsubara formalism \cite{abrikosov:1963,mahan:1990}, we establish the equivalence of our results with the Kubo formula {\em in the ladder approximation}. An application to spin-orbit torque (SOT) theory in the two-dimensional electron gas (2DEG) demonstrates the distinct advantages of our approach, and uncover a new instance of intrinsic response as a Fermi surface property, reminiscent of the classical result obtained by Haldane in the case of the AHE \cite{haldane:2004}. Finally, we outline some implications for the theory of the recently observed orbital edge accumulation \cite{lyalin:2023,idrobo:2024} and orbital torque \cite[\& references therein]{sala:2022}, and we conclude by mentioning some future extensions. 

{\em General formalism  --} Our main new result is a general expression for the interband part of the density matrix {\em i.e.}, \smash{$\hat{\rho}_{inter}^{\scriptscriptstyle (1)} = \sum_{n \neq m} \hat{\rho}_{n m}^{\scriptscriptstyle (1)}$}, in linear response to \smash{${\bm E}$}, with
\begin{align} \label{eq:rho_tran_lin_version_B}
    \!\!\!\hat{\rho}_{n m}^{\scriptscriptstyle (1)} &= - e \hbar \frac{f_{n}^{\scriptscriptstyle (eq)}- f_{m}^{\scriptscriptstyle (eq)} }{\varepsilon_n - \varepsilon_m}    {\bm E} \cdot \hat{\mathcal{\bm A}}_{n m} - \hbar \partial_{\bm x} \frac{f_n^{\scriptscriptstyle (1)}+f_m^{\scriptscriptstyle (1)}}{2} \cdot \hat{\mathcal{\bm A}}_{n m}   \nonumber  \\
    &\!\!\!\!\!\!\!  - \frac{{\bf i} \hbar}{2 (\varepsilon_n - \varepsilon_m)} \frac{2 \pi n_i v_0^2}{\hbar} \sum_{q} \int \!\! \frac{d^d \underaccent{\tilde}{p}}{(2 \pi \hbar)^d}  \hat{P}_n \underaccent{\tilde}{\hat{P}}_q \hat{P}_m  \times \cdots  \\
    &\!\!\!\!\!\!\!\!\!\!\! \cdots \times \left[ \delta \left( \underaccent{\tilde}{\varepsilon}_{q} \! - \! \varepsilon_{n} \right) \left( \underaccent{\tilde}{f}_q^{\scriptscriptstyle (1)} \!\! - \!\! f_n^{\scriptscriptstyle (1)} \right) + \delta \left( \underaccent{\tilde}{\varepsilon}_{q} \! - \! \varepsilon_{m} \right) \left( \underaccent{\tilde}{f}_q^{\scriptscriptstyle (1)} \!\! - \!\! f_m^{\scriptscriptstyle (1)} \right)  \right] , \nonumber
\end{align}
while the intraband part reads \smash{$\hat{\rho}_{intra}^{\scriptscriptstyle (1)} = \sum_{n} f^{\scriptscriptstyle (1)}_{n} \hat{P}_n$} \footnote{This form of \smash{$\hat{\rho}^{\scriptscriptstyle (1)}$} is strictly valid for non degenerate bands. In the case of extended degeneracy, which is most generically realized in the form of Kramer's degeneracy in time reversal and inversion symmetric systems, it remains valid if the spin-orbit interaction can be safely neglected. Otherwise, a traceless matrix part shall also be considered. In addition, in Eq.(1) we omit principal value terms which are usually interpreted as disorder-induced renormalization terms of the Hamiltonian. Both  can be found in \cite{supp}.}  and is controlled by a system of linear BEs
\begin{equation} \label{eq:bolt_lin}
    \partial_t f_n^{\scriptscriptstyle (1)} + \partial_{\bm p} \varepsilon_n \cdot \partial_{\bm x} f_n^{\scriptscriptstyle (1)} + (e {\bm E}) \cdot \partial_{\bm p} \varepsilon_n \partial_{\varepsilon} f_{n}^{\scriptscriptstyle (eq)}  =  C_n[{\bm f}^{\scriptscriptstyle (1)}] .
\end{equation} 
To be more precise, \smash{$\hat{\rho}^{\scriptscriptstyle (1)}({\bm x}, t ; {\bm p}) \equiv \hat{\rho}^{\scriptscriptstyle (1)} = \hat{\rho}^{\scriptscriptstyle (1)}_{intra} +  \hat{\rho}^{\scriptscriptstyle (1)}_{inter}$} as determined by Eqs.(\ref{eq:rho_tran_lin_version_B}-\ref{eq:bolt_lin}) is the change in the gauge invariant Wigner function {\em i.e.}, in the gauge invariant Wigner transform of the density matrix \cite{serima:1986}. Once \smash{$\hat{\rho}^{\scriptscriptstyle (1)}$} is obtained, the linear response for any single particle observable $\smash{{\hat O}}$, in terms of its local density variation $\smash{o^{\scriptscriptstyle (1)}({\bm x}, t)}$, is obtained from its Weyl symbol $\smash{{\hat o}({\bm x}, {\bm p})}$ as \smash{$o^{\scriptscriptstyle (1)}({\bm x}, t) = \int \frac{d^d p}{(2 \pi \hbar)^d} \ \mathrm{Tr} [ {\hat \rho}^{\scriptscriptstyle (1)}({\bm x}, t; {\bm p}) \ {\hat o}({\bm x}, {\bm p}) ]$.} Equations  (\ref{eq:rho_tran_lin_version_B}-\ref{eq:bolt_lin}) determine the linear response of a non interacting system of electrons with charge $e(<0)$, specified by a multiband low energy Hamiltonian matrix {\em i.e.}, \smash{${\hat h} = {\hat h}({\bm p})$}, with ${\bm p}$ the {\em kinetic} crystal pseudo momentum. This matrix is acting on the Hilbert space of dimension $N$ of the internal degrees of freedom (spin, orbital), and it may equally correspond to a ``toy'' model or to a sophisticated Wannier tight-binding Hamiltonian \cite{foulkes:1989,goringe:1997,garrity:2021}. It has in general $M \leq N$ {\em distinct} eigenvalues, the band energies, introduced as \smash{$\varepsilon_n \equiv \varepsilon_n ({\bm p})$.} The projectors into the corresponding eigenspaces, introduced as \smash{$\hat{P}_n \equiv \hat{P}_n ({\bm p})$,} read \cite{graf:2021} \smash{$\hat{P}_n = \prod_{m} \frac{\hat{h} - \varepsilon_m \hat{\gamma}_0}{\varepsilon_n - \varepsilon_m}$,} with $\hat{\gamma}_0$ the identity matrix. We further define \smash{$\hat{\mathcal{\bm A}}_{n m} = {\bf i} \hat{P}_n \partial_{\bm p} \hat{P}_m$,}
which appears in the first line of Eq.(\ref{eq:rho_tran_lin_version_B}). In the special case of two non degenerate bands, it is immediate to verify that we have \smash{$(\hat{\mathcal{\bm A}}_{n m})^i = {\mathcal{A}}^i_{n m} | n \rangle  \langle m |$, with ${\mathcal{A}}^i_{n m} = {\bf i} \langle n | \partial_{p^i} m \rangle$} a previously introduced abelian interband Berry connection \cite{culcer:2017,wang:2023}. Hence, we recognize \smash{$\hat{\mathcal{\bm A}}_{n m}$} as a generalized {\em non abelian interband  Berry connection}, defined in a manifestly $SU(N)$ gauge invariant way.  Also appearing in Eqs.(\ref{eq:rho_tran_lin_version_B}-\ref{eq:bolt_lin}), are the linear changes in the intraband distribution functions {\em i.e.}, \smash{$f_n^{\scriptscriptstyle (1)} \equiv f_n^{\scriptscriptstyle (1)} ({\bm x}, t ; {\bm p})$}, reducing in equilibrium to the Fermi-Dirac distribution function \smash{$f_n^{\scriptscriptstyle (eq)} \equiv n_{\scriptscriptstyle FD} (\varepsilon_n)$}. In the present work, we purposely limit our consideration of disorder to the most simple and generic model {\em i.e.}, a dilute random distribution of scalar point scattering potentials treated at the level of the self-consistent second Born approximation after configuration averaging. It is parameterized by the density $n_i$ of defects and by a momentum independent scattering amplitude $v_0$. The collision integral appearing on the right-hand-side of the BE (\ref{eq:bolt_lin}) reads
\begin{equation}
     \!\!\!\! C_n[{\bm f}^{\scriptscriptstyle (1)}] \!   = \!\! \sum_{q} \!\! \int \!\!\! \frac{d^d \underaccent{\tilde}{p}}{(2 \pi \hbar)^d} \delta \left( \underaccent{\tilde}{\varepsilon}_{q} - \varepsilon_{n} \right)
    W_{n q}  \left( \underaccent{\tilde}{f}_q^{\scriptscriptstyle (1)} - f_n^{\scriptscriptstyle (1)} \right) \! ,
    \label{eq:collision_integral}
\end{equation}
with the scattering kernel given by
\begin{equation}
    W_{n q}\equiv W_{n q}({\bm p},  \underaccent{\tilde}{\bm p}) = \frac{2 \pi n_i v_0^2}{\hbar w_n}  {\rm Tr} \left[ 
    \hat{P}_n \underaccent{\tilde}{\hat{P}}_q \right] ,
    \label{eq:scattering_kernel}
\end{equation}
with $w_n \geq 1$ being the degree of degeneracy of band $n$, and with a tilde under-accent indicating throughout this Letter that a certain quantity depends on  \smash{momentum $\underaccent{\tilde}{\bm p}$.} After introducing  \smash{$\tau \propto \hbar/(n_i v_0^2)$} as the relaxation time scale, we see that \smash{$\xi = 1 / \tau$} defines a natural scaling parameter for the disorder strength, while $\hbar$ is of course the formal expansion parameter for the quantum corrections to the classical kinetics \cite[\& references therein]{hidaka:2022,valet:2023}. 

While the commonly considered decomposition of the Kubo-Streda type formulas into Fermi sea and Fermi surface terms \cite{smrcka:1977,bonbien:2020} is not physically significant in itself \cite{haldane:2004}, the quantum kinetic approach leads naturally to a unique and gauge invariant split at the Wigner function level {\em i.e.}, \smash{$\hat{\rho}^{\scriptscriptstyle (1)} = \hat{\rho}^{\scriptscriptstyle (1)}_{intra} +  \hat{\rho}^{\scriptscriptstyle (1)}_{inter}$}. Its physical meaning is further illuminated by Eq.(\ref{eq:rho_tran_lin_version_B}), establishing the interband response components as {\em dependent quantities locally determined} by the quantum geometry and the solutions of the BE (\ref{eq:bolt_lin}). Hence, the quantum kinetic formulation provides the key new physical insight that interband contributions to observable densities are {\em not} quantities susceptible to  advective or diffusive transport in real space. Looking at the right-hand side (RHS) of Eq.(\ref{eq:rho_tran_lin_version_B}), we recognize the first term as of purely quantum geometric origin and of order $\hbar \xi^0$, which yields for instance the intrinsic contribution to the AHE \cite{jungwirth:2002,nagaosa:2010} controlled by the Berry curvature in reciprocal space. If we look now at the third term on the RHS of our Eq.(\ref{eq:rho_tran_lin_version_B}), since \smash{$f_n^{\scriptscriptstyle (1)} \sim 1/\xi$}, as it can be immediately verified by direct inspection of Eqs.(\ref{eq:bolt_lin}-\ref{eq:scattering_kernel}), we see that it is also of order $\hbar \xi^0$ due to the compensating factor \smash{$(n_iv_0^2)/\hbar\propto \xi$} in front of the momentum integral. Hence, even if this contribution is disorder mediated, it is ultimately independent of its strength. As demonstrated in the next section, this corresponds to the vertex correction in the ladder approximation to the interband part of the Kubo formula, for which we provide here an {\em explicit and compact expression}, a substantial advantage compared to the Kubo formula approach for concrete calculations. Moreover, as previously mentioned, even recent works rely most often on a simplified version of the Kubo-Streda formula at the CBA level, with the leading disorder correction to the interband response found to be of order $\hbar \xi$ \cite{li:2015}, which is qualitatively wrong and totally misses the true leading effect of disorder. The second term on the RHS of Eq.(\ref{eq:rho_tran_lin_version_B}) indicates that mesoscopic gradients in spatially inhomogeneous solutions of the BE (\ref{eq:bolt_lin}) are also a source of interband quantum coherences, for which we provide here an explicit expression. To the best of our knowledge, this is an entirely new finding. Such gradients typically appear near boundaries and interfaces on the length scale of the electron mean-free-path \smash{$l_e \propto 1/\xi$} \cite{fuchs:1938,sondheimer:1952,camley:1989}. This opens a new area of research, since one can envision to engineer such gradients using nanofabrication techniques. Since \smash{$\hbar \partial_{\bm x} f_n^{\scriptscriptstyle (1)} \sim (\hbar / l_e) f_n^{\scriptscriptstyle (1)} \sim \hbar \xi^0$,} this is of the same order as the  other contributions. Regarding the intraband response, it shall also be noted that there is {\em no} Berry curvature corrections to the BE (\ref{eq:bolt_lin}), at the leading order \smash{$ f_n^{\scriptscriptstyle (1)} \sim 1/\xi$.} There are however sub-leading terms at order $\hbar \xi^0$ {\em i.e.}, in particular anomalous velocity, intrinsic side-jump and skew-scattering, previously found with the semiclassical approach \cite{sinitsyn:2007,sinitsyn:2008,xiao:2017}. This is beyond the scope of the present letter, since a complete account requires to go beyond the ladder approximation \cite{ado:2016}, and will be provided within the same general framework in a future work. 

{\em Correspondence with the Kubo formula--} It is well established that the intraband part of the linear response controlled by the BE (\ref{eq:bolt_lin}), in the constant and homogeneous case, is identical to the intraband part of the Kubo formula with the vertex correction in the ladder approximation  \cite{abrikosov:1963,kim:2019b}. In this section, we establish a similar equivalence for the interband response as given by Eq.(\ref{eq:rho_tran_lin_version_B}). According to the Kubo formula at finite temperature in the constant and translationally invariant limits \cite{abrikosov:1963,mahan:1990}, the expectation value  of an observable \smash{${\hat o}(\bm{p}) \equiv {\hat o}$} reads \smash{$o^{\scriptscriptstyle (1)} = - e E^i \int_{p} \lim_{\omega\rightarrow 0} R^i (\omega)/({\bf i} \omega)$,} with \smash{$\int_{p} \equiv \intvol{{p}}$} and with \smash{$R^i ({\bm p}, \omega) \equiv R^i (\omega)$} being the analytical continuation (AC) of the Matsubara response function {\em i.e.},
\smash{$R^i (\omega ) ={\cal R}^i (\omega_{\nu}\rightarrow - {\bf i}(\hbar \omega + {\bf i} 0^+))$}, with
\begin{equation}
{\cal R}^i (\omega_{\nu}) =  \sum_{n,m,\varepsilon_{\mu}}{\rm Tr}\left[\hat{P}_n {\hat o} \hat{P}_m \hat{\mathcal{G}}(\varepsilon_{\mu} +\omega_{\nu}) {\hat J}^i \hat{\mathcal{G}}(\varepsilon_{\mu})\right],
\label{sec_LRT_5}
\end{equation}
with \smash{$\sum_{\epsilon_{\mu}} \equiv (1/\beta) \sum_{\varepsilon_{\mu}}$} and \smash{$\hat{\mathcal{G}}({\bm p}, \varepsilon_{\mu}) \equiv \hat{\mathcal{G}}(\varepsilon_{\mu})$} the matrix-valued Matsubara Green's function, while \smash{$\varepsilon_{\mu}$} and \smash{$\omega_{\nu}$} are fermionic and bosonic frequencies and \smash{${\hat J}^i (\bm{p}) \equiv {\hat J}^i$} is the disorder-dressed current vertex. The Bethe-Salpeter equation for this vertex, in the ladder approximation, is 
\begin{equation}
\label{vertex_salpeter}
{\hat J}^i = {\hat j}^i
+ \diso \int_{\underaccent{\tilde}{p}}
\underaccent{\tilde}{\hat{\mathcal{G}}}(\epsilon_{\mu} +\omega_{\nu}) \underaccent{\tilde}{\hat J}^i \underaccent{\tilde}{\hat{\mathcal{G}}}(\epsilon_{\mu}) ,
\end{equation}
with \smash{${\hat j}^i = \partial_{p^i} \hat{h}$} the bare current vertex. From the corresponding expression of $o^{\scriptscriptstyle (1)}$ in terms of the density matrix, we immediately establish the correspondance
\begin{equation}
    \!\! {\hat \rho}^{\scriptscriptstyle (1)}_{nm} =  - e  E^i \!\lim_{\omega\rightarrow 0} \frac{\left[ \sum_{\epsilon_{\mu}} \!\!\hat{P}_n \hat{\mathcal{G}}(\epsilon_{\mu} +\omega_{\nu}) {\hat J}^i \hat{\mathcal{G}}(\epsilon_{\mu})\hat{P}_m  \right]_{\scriptscriptstyle AC}}{{\bf i}\omega} , \label{eq:inter_density_matrix_diagram}
\end{equation}
in which the AC is being performed before taking the \smash{$\omega\rightarrow 0$} limit. The Matsubara Green's function \smash{$\hat{\mathcal{G}}(\epsilon_{\mu})$}, in presence of disorder, can be obtained from the retarded (advanced) Green's functions \smash{$\hat{g}^{R(A)}({\bm p}, \varepsilon) \equiv \hat{g}^{R(A)}$,} which corresponds to its analytical continuation in the upper (lower) complex plane. Introducing the decomposition \smash{$\hat{g}^{\scriptscriptstyle R(A)} = \sum_n g^{\scriptscriptstyle R(A)}_n \hat{P}_n + \sum_{n \neq m} \hat{P}_n \hat{g}^{\scriptscriptstyle R(A)}_{n m} \hat{P}_m$}, and at the level of the self-consistent Born approximation, we have \smash{
$g^{\scriptscriptstyle R(A)}_n = [ \varepsilon - \varepsilon_{n} \substack{\scriptscriptstyle + \\ \scriptscriptstyle (-)} {\bf i}\hbar / ( 2\tau_n) ]^{-1}$}, with \smash{$\frac{1}{\tau_{n}} =  \int_{\underaccent{\tilde}{p}} \sum_m \delta (\varepsilon_{n} - \underaccent{\tilde}{\varepsilon}_{m} ) W_{n m}$}, while
\begin{equation}
\!\!\! \hat{g}^{\scriptscriptstyle R(A)}_{n m} = \substack{\scriptscriptstyle + \\ \scriptscriptstyle (-)} 2 {\bf i} \pi \diso g^{\scriptscriptstyle R(A)}_n g^{\scriptscriptstyle R(A)}_m \sum_q \!\!
\int_{\underaccent{\tilde}{p}} \! 
\delta (\varepsilon - \underaccent{\tilde}{\varepsilon}_{m} ) \hat{P}_n \underaccent{\tilde}{\hat{P}}_q \hat{P}_m .
\label{eq:g_5}
\end{equation}
We can also write \smash{${\hat J}^i = \sum_n J^i_n \hat{P}_n + \sum_{n \neq m} \hat{P}_n {\hat J}^i_{n m} \hat{P}_m$}. At leading order \mbox{in $\xi$,} we then obtain the band projected part of Eq.(\ref{vertex_salpeter}) as
\begin{equation}
J_n^i = j_n^i +\theta_{\mu \nu}
\int_{\underaccent{\tilde}{p}} \sum_m \delta (\varepsilon - \underaccent{\tilde}{\varepsilon}_{m}) \underaccent{\tilde}{\tau}_{m} \underaccent{\tilde}J^i_m W_{nm} ,\label{eq:band_vertex_longitudinal}
\end{equation}
and its interband components as
\begin{equation}
\!\!\! {\hat J}^i_{n m} \! = \! {\hat j}^i_{n m} + \frac{2\pi n_i v_0^2}{\hbar} \theta_{\mu \nu}
\int_{\underaccent{\tilde}{p}}  \sum_{q} 
\delta (\varepsilon -\underaccent{\tilde}{\varepsilon}_{q} ) \underaccent{\tilde}{\tau}_{q}
 \underaccent{\tilde}{J}_q^i \hat{P}_n \underaccent{\tilde}{\hat{P}}_q \hat{P}_m ,
\label{eq:transverse_vertex}
\end{equation}
with \smash{$\theta_{\mu \nu}\equiv \theta (-\epsilon_{\mu}(\epsilon_{\mu}+\omega_{\nu}))$,} by using the analytic structure of the Matsubara Green's function \cite{supp}. From Eq.(\ref{eq:inter_density_matrix_diagram}) with $n=m$, one obtains at the same leading order \smash{$f_{n}^{\scriptscriptstyle (1)} = -\partial_{\varepsilon} f_{n}^{\scriptscriptstyle (eq)}
 \tau_{n} \sum_i (e E^i) J^i_n$},
which confirms the expected equivalence between Eq.(\ref{eq:band_vertex_longitudinal}) and Eq.(\ref{eq:bolt_lin}), and provides an explicit expression for \smash{$J^i_n$} in terms of \smash{$f_{n}^{\scriptscriptstyle (1)}$.} Then, going back to Eq.(\ref{eq:inter_density_matrix_diagram}), it is clear that the interband components of the density matrix are either contributed by the interband components of the current vertex or by those of the two Green's functions. Then, the first term on the RHS of Eq.(\ref{eq:transverse_vertex}) gives the first term on the RHS of Eq.(\ref{eq:rho_tran_lin_version_B}). As for the third term on the RHS of Eq.(\ref{eq:rho_tran_lin_version_B}), the second term of Eq.(\ref{eq:transverse_vertex})) yields the terms proportional to \smash{$\underaccent{\tilde}{f}^{\scriptscriptstyle (1)}_q$,} whereas the interband components of the Green's functions, as derived from Eq.(\ref{eq:g_5}), lead to the terms proportional to \smash{$f^{\scriptscriptstyle (1)}_n$} and \smash{$f^{\scriptscriptstyle (1)}_m$ \cite{supp}.} This completes the demonstration of the equivalence between our formalism and the Kubo formula in the ladder approximation, in the constant and homogeneous limits. One needs of course to add that our formalism also validly captures the part of the linear responses {\em slowly} varying in time and space at the same leading orders in the semiclassical and disorder expansions, in the case of finite mesoscopic systems. This is much more challenging to achieve with fully quantum calculations. 

{\em Pedagogical application to the SOT in the 2DEG--} As a pedagogical example, and as a way to demonstrate the advantages of our kinetic formulation when performing concrete calculations beyond the CBA, we address now the microscopic theory of the SOT for a weakly disordered 2DEG in presence of exchange splitting induced by some classical magnetization vector, a topic previously investigated with the Kubo formula \cite{ado:2017,veneri:2022}. The model Hamiltonian is
\begin{equation} \label{eq:ham_init}
    \hat{h}({\bm p}) = \varepsilon_0 ({\bm p}) \hat{\sigma}_0 + J_{e} {\bm m} \cdot \hat{\bm \sigma} + {\bm b}_{\scriptscriptstyle SO} ({\bm p}) \cdot \hat{\bm \sigma} ,
\end{equation}
with \smash{${\bm p} \equiv (p^x, p^y)$}, \smash{$\varepsilon_0$} some paramagnetic band dispersion, \smash{$J_e > 0$} the exchange constant, \smash{${\bm m}$} a classical unit vector aligned with the magnetization and \smash{${\bm b}_{\scriptscriptstyle SO}$} some momentum dependent SO field with \smash{$b_{\scriptscriptstyle SO} \ll J_e$.} The operator \smash{$\hat{\bm \sigma}$} is the vector of Pauli matrices and corresponds here to the electron spin operator, while \smash{$\hat{\sigma}_0$} is the identity matrix in spin space. The Hamiltonian (\ref{eq:ham_init}) is a generalization, to arbitrary band and SO field dispersions, of the model considered in \cite{ado:2017}. If we neglect dissipation \cite{gilbert:1955,gilbert:2004}, the dynamics of ${\bm m}$ is controlled by the Landau-Lifschitz equation \cite{brown:1963,andreev:1980} {\em i.e.}, \smash{$(M_{\scriptscriptstyle S} / \gamma) \partial_t {\bm m} = {\bm m} \times ( \delta  \mathcal{F} / \delta {\bm m} ) \equiv {\bm T}$}, with \smash{$\gamma(<0)$} the gyromagnetic ratio, \smash{$M_{\scriptscriptstyle S}$} the saturation magnetization, $\mathcal{F}$ the thermodynamic potential and ${\bm T}$ the torque density. The SOT, in linear response to a quasi uniform and constant electric field, is then given by \smash{${\bm T}^{ (1)} = {\bm m} \times ( \delta  \mathcal{F}^{\scriptscriptstyle (1)} / \delta {\bm m} )$}, with \smash{$\mathcal{F}^{\scriptscriptstyle (1)}$} the linear change in $\mathcal{F}$. By application of the Hellmann-Feynman theorem \cite{hellmann:1933,feynman:1939} at finite temperature \cite{pons:2020}, and assuming a spatially uniform magnetization, we then obtain \smash{${\bm T}^{\scriptscriptstyle (1)} = {\bm m} \times \langle \delta \hat{h} / \delta {\bm m} \rangle^{\scriptscriptstyle (1)} = J_e {\bm m} \times \langle \hat{\bm \sigma} \rangle^{\scriptscriptstyle (1)}$.} Hence we see that the SOT is controlled by the component of the expectation value of the spin density perpendicular to the classical magnetization, in linear response to the field. In the case of an SO field breaking inversion symmetry, one expects both intraband and interband contributions to the SOT \cite{manchon:2019}. Here, we uniquely focus on the interband contribution, for which a previous Kubo calculation has established in some special cases the cancellation of the intrinsic SOT by the vertex correction \cite{ado:2017}. 

We start by deriving anew the intrinsic SOT {\em i.e.}, the contribution of purely quantum geometric origin. The band energies and associated projectors for the Hamiltonian (\ref{eq:ham_init}) are trivially obtained as \smash{$\varepsilon_\sigma = \varepsilon^0 + \sigma \Delta$} and \smash{$\hat{P}_\sigma = \frac{1}{2} \left( \hat{\sigma}_0 + \sigma {\bm u} \cdot \hat{\bm \sigma} \right)$,} respectively ; in which we have introduced \smash{$\sigma \in \lbrace +1, -1 \rbrace \equiv \lbrace \uparrow, \downarrow \rbrace$,} with \smash{$\Delta \approx J_{e} ( 1 +  {\bm m} \cdot \underaccent{\bar}{\bm b}_{\scriptscriptstyle SO} )$} and \smash{${\bm u} \approx {\bm m}  + {\bm m} \times ( \underaccent{\bar}{\bm b}_{\scriptscriptstyle SO} \times {\bm m} )$,} in which the almost equal symbol is used from now on to indicate quantities approximated at leading order in the SO field normalized by the exchange energy {\em i.e.}, \smash{$\underaccent{\bar}{\bm b}_{\scriptscriptstyle SO} = {\bm b}_{\scriptscriptstyle SO} / J_e$}, which is a natural small parameter of the problem. The interband Berry connection reads \smash{$ \hat{\mathcal{A}}^i_{\sigma \underaccent{\tilde}{\sigma}} \ \approx \ \frac{{\bf i} \underaccent{\tilde}{\sigma}}{4} \partial_{p^i} \underaccent{\bar}{\bm b}_{\perp} \cdot \hat{\bm \sigma} \ + \ \frac{1}{4} ({\bm m} \times \partial_{p^i} \underaccent{\bar}{\bm b}_{\perp} ) \cdot \hat{\bm \sigma}$,} with \smash{$\underaccent{\bar}{\bm b}_{\perp} = {\bm m} \times ( \underaccent{\bar}{\bm b}_{\scriptscriptstyle SO} \times {\bm m} )$.} The intrinsic contribution to the SOT is then readily obtained from the trace of the first term on the RHS of Eq.(\ref{eq:rho_tran_lin_version_B}) with the spin operator, followed by momentum integration over the 2D Brillouin zone (BZ), and it reads 
\begin{equation} \label{eq:int_sea}
    {\bm T}^{\scriptscriptstyle (1)}_{int} \approx  \frac{\hbar}{2} \int\displaylimits_{BZ} \!\! \frac{d^2 \underaccent{\tilde}{p}}{(2 \pi \hbar)^2} \left( f^{\scriptscriptstyle (eq)}_\uparrow - f^{\scriptscriptstyle (eq)}_\downarrow \right) ( e {\bm E} )  \cdot \partial_{\bm p}  \underaccent{\bar}{\bm b}_{\perp} .
\end{equation}
Before considering the disorder correction, it is relevant to first recognize that \smash{${\bm E}  \cdot \partial_{\bm p}  \underaccent{\bar}{b}_{\perp}^i = \partial_{\bm p} \cdot ( \underaccent{\bar}{b}_{\perp}^i {\bm E} )$}, since under the additional assumption of the system to be degenerate, an application of the divergence theorem to the RHS of Eq.(\ref{eq:int_sea}) then yields
\begin{equation} \label{eq:int_surf}
    {\bm T}^{\scriptscriptstyle (1)}_{int} \approx  \frac{\hbar}{2} \sum_{\sigma} \sigma \int\displaylimits_{FS_\sigma} \!\! \frac{dl_p}{(2 \pi \hbar)^2} \ \frac{( e {\bm E} ) \cdot \partial_{\bm p} \varepsilon_0}{|\partial_{\bm p} \varepsilon_0|}  \underaccent{\bar}{\bm b}_{\perp} ,
\end{equation}
with $dl_p$ being the infinitesimal line element along the 1D Fermi surface for band $\sigma$ in the absence of SO interaction {\em i.e.}, \smash{$FS_\sigma$}. One shall add that if the Fermi surface intersects the BZ boundary, the line integrals over the portions of this boundary resulting from the application of the divergence theorem do {\em not} contribute to the RHS of Eq.(\ref{eq:int_sea}). Indeed, since the 2D BZ is by definition a primitive unit cell of the reciprocal space over which \smash{$\hat{h}$} is periodic, such contributions will cancel out by pairs, from pairs of line segments on opposite sides of the BZ boundary, one being the translation of the other by a primitive reciprocal lattice vector. Hence, we have established that under the stated hypothesis, the intrinsic SOT, which according to Eq.(\ref{eq:int_sea}) would be classified as a Fermi sea contribution, can equally be expressed as a Fermi surface one according to Eq.(\ref{eq:int_surf}). This exactly parallels the classical result obtained by Haldane in the case of the non quantized part of the AHE conductivity \cite{haldane:2004}, and is a further illustration that the decomposition of the linear response into Fermi sea and Fermi surface terms emphasized by the Kubo-Streda formula is neither unique nor particularly relevant.  

Having established this non trivial result, we demonstrate now how our formalism delivers the disorder correction with striking efficiency. From the third term on the RHS of our Eq.(\ref{eq:rho_tran_lin_version_B}), we readily obtain the vertex correction to the interband Wigner function as
\begin{equation} \label{eq:rho_dis}
    \!\!\!\! \hat{\rho}^{\scriptscriptstyle (1)}_{dis} \! \approx \! \frac{\pi n_i v_0^2}{2 J_e}  {\bm m} \times \!\sum_{\sigma, \underaccent{\tilde}{\sigma}}  \underaccent{\tilde}{\sigma} \!\!\!\!\int\displaylimits_{{FS}_{\scriptscriptstyle \underaccent{\tilde}{\sigma}}} \!\!\!\!\! \frac{dl_{\underaccent{\tilde}{p}}}{(2 \pi \hbar)^2} \! \frac{ f_\sigma^{\scriptscriptstyle (1)} \! - \! \underaccent{\tilde}f_{\underaccent{\tilde}{\sigma}}^{\scriptscriptstyle (1)}}{| \partial_{\underaccent{\tilde}{\bm p}} \underaccent{\tilde}{\varepsilon}_{0} |}  ( \underaccent{\bar}{\bm b}_{\perp} \! -  \underaccent{\tilde}{\underaccent{\bar}{\bm b}}_{\perp} ) \cdot  \hat{\bm \sigma} ,
\end{equation}
with the distribution function to be derived without SO interaction. In this limit, the scattering Kernel \smash{$W_{\sigma, \underaccent{\tilde}{\sigma}}$} becomes diagonal and the BEs for the two spin bands decouple. At this point, we further specialize the problem to a parabolic 2DEG  as in \cite{ado:2017} {\em i.e.}, to \smash{$\varepsilon_0 \equiv p^2 / (2 m^\star)$} with \smash{$m^\star$} the effective mass. One can then readily solve the BEs with \smash{$f_{\sigma}^{\scriptscriptstyle (1)} = \delta \left({\varepsilon}_\sigma - \varepsilon_{\scriptscriptstyle F} \right) ( \tau_F / m^\star ) ( e {\bf E} ) \cdot {\bm p}$}, with the relaxation time and the density of states at the Fermi level respectively given by \smash{$1/\tau_F = (2 \pi n_i v_0^2 D_F)/\hbar$} and \smash{$D_F = m^\star / (2 \pi \hbar^2)$.} Substituting this into Eq.(\ref{eq:rho_dis}), and computing the associated spin density, leads to
\begin{equation} \label{eq:int_dis}
    {\bm T}^{\scriptscriptstyle (1)}_{dis} \approx - \frac{\hbar}{2} \sum_{\sigma} \sigma \int\displaylimits_{FS_\sigma} \!\! \frac{dl_p}{(2 \pi \hbar)^2} \ ( e {\bm E} ) \cdot\frac{\bm p}{p} \ \underaccent{\bar}{\bm b}_{\perp} ,
\end{equation}
exactly canceling the intrinsic torque as given by Eq.(\ref{eq:int_surf}). This generalizes, to arbitrary weak SO interaction, the central result obtained in \cite{ado:2017} with considerable more efforts while relying on a diagrammatic expansion of the Kubo formula in the ladder approximation, for an SO interaction restricted to Rashba or Dresselhaus types. 

{\em Implications for the theory of orbital effects --} Regarding the theory of out-of-equilibrium orbital angular momentum (OAM) in time reversal and inversion symmetric metals with weak SO interaction, of high current interest in connection with recent experiments \cite{sala:2022,lyalin:2023,idrobo:2024}, most Kubo formula based studies have focused on the intrinsic OHE conductivity \cite[\& references therein]{pezo:2022}. The intrinsic OHE current can be straightforwardly computed from the first term on the RHS of our Eq.(\ref{eq:rho_tran_lin_version_B}). Beyond the possible fragility of this current under disorder corrections \cite{tang:2024}, which can be very effectively assessed from the third term on the RHS of our Eq.(\ref{eq:rho_tran_lin_version_B}), and its debatable physical meaning as a non conserved quantity \cite{bernevig:2005}, we want to point out the importance of the gradient induced quantum coherence, as newly uncovered by the second term on the RHS of our Eq.(\ref{eq:rho_tran_lin_version_B}). It can be shown \cite{valet:2025b} that this is a source of OAM edge accumulation, only partially correlated to the OHE conductivity through the Berry connection, and which may be an important explanatory factor to the previously cited experimental observations.

{\em Conclusions --} We provide a quantum kinetic theory of the linear response to  an electric field slowly varying in time and space, for multiband electron systems in presence of weak scalar disorder. While our theory constitutes a formal and computational alternative to the Kubo formula in the stated context, with distinct advantages for finite systems and in going beyond the CBA, we have proven the equivalence of the two formulations at the ladder approximation level in the infinite and uniform case. The natural and unique decomposition of the linear response in our formalism, in term of an intraband response controlled by a transport equation in phase space, and interband coherences as local dependent quantities, provides new physical insights in the quantum structure of the linear response which are totally obfuscated in the Kubo formula. We have uncovered for the first time that spatial gradients in the intraband response on mesoscopic scales, which are ubiquitous in finite conductors, constitute a source of interband coherence. This opens a new research area as it suggests new ways to locally modulate current induced responses {\em e.g.}, spin or orbital accumulation, by controlling the real space geometry or the boundary scattering properties.  We have applied our approach to SOT theory for a simple model of 2DEG, recovering and generalizing with striking efficiency a major vertex correction, while uncovering a new instance of intrinsic response as a Fermi surface property. We also suggest possible implications for the theory of orbital effects, a subject of high current interest both from a fundamental and applied standpoint. Beyond analytical toy models, our formalism is equally applicable to realistic material models, up to ab-initio generated tight-binding Hamiltonian, through numerical resolution of the Boltzmann equation. We are already undertaking such studies in relation to SOT in van der Waals magnets with $C_3 v$ symmetry \cite{valet:2025}, another subject of high current interest \cite{ovalle:2023}. Finally, we shall stress that the simple disorder model considered in this Letter can be easily extended, for instance to include SO scattering or contributions from higher order diagrams beyond the ladder approximation, and that we also intend to address in a systematic manner the sub-leading contributions to the intraband response in a future work.                  

\begin{acknowledgments}
We thank Albert Fert, Henri Jaffr\`es, Vincent Cros, Mairbek Chshiev, Sergey Nikolaev, Jing Li, Libor Voj\'a\v{c}ek, Gerrit E.W. Bauer, Fr\'ed\'eric Pi\'echon and Xavier Waintal for fruitful discussions. One of the authors (TV) acknowledges funding from the Laboratoire Albert Fert (UMR137)
CNRS, Thales, Université Paris-Saclay, France. 
\end{acknowledgments}

\bibliography{berry}

\end{document}